\documentclass[12pt,a4paper]{article}
\usepackage{amssymb,amsfonts,amsmath}

\begin{document}

\title{Can there be given any meaning to contextuality without incompatibility?}

\author{Andrei Khrennikov\\ 
Linnaeus University, International Center for Mathematical Modeling\\  in Physics and Cognitive Sciences
 V\"axj\"o, SE-351 95, Sweden}

\maketitle

\abstract{Our aim is to compare  the fundamental notions of quantum physics -  contextuality vs. incompatibility. 
One has to distinguish two different notions of contextuality,  {\it Bohr-contextuality} and {\it Bell-contextuality}.
The latter is defined operationally via violation of noncontextuality (Bell type) inequalities. This sort of contextuality will be compared with  incompatibility. It is easy to show that, for quantum observables, there is {\it no contextuality without incompatibility.}
The natural question arises:  What is contextuality without  incompatibility? (What is ``dry-residue''?)
Generally this is the very complex question. We concentrated on contextuality for four quantum observables.  We shown that in the CHSH-scenarios (for ``natural quantum observables'') {\it contextuality is reduced to incompatibility.} However, generally contextuality without incompatibility may have  some physical 
content. We found a mathematical constraint extracting the contextuality component from incompatibility. However, the physical meaning of this constraint is not clear.  In appendix 1, we briefly discuss another sort of contextuality based on the Bohr's complementarity principle which is treated as  the {\it contextuality-incompatibility principle}. Bohr-contextuality plays the crucial role  in quantum foundations.  Incompatibility is, in fact, a consequence of Bohr-contextuality. Finally, we remark 
that outside of physics, e.g., in cognitive  psychology and  decision making  
Bell-contextuality cleaned of incompatibility can play the  important role. } 

keywords: Bell-contextuality; Bohr-contextuality; incompatibility; complementarity principle;  
joint probability distribution; noncontextual inequalities; product of commutators. 
   
\section{Introduction}

Contextuality formalized in the form of violation of  noncontextuality inequalities, {\it Bell-contextuality} \cite{Bell1,Bell2}, is a hot topic in quantum physics (see, e.g., \cite{AD0,AD} and references herein). Unfortunately, it is typically presented in the  mathematical framework and its physical meaning is unclear. 

We stress that, in fact, one has to distinguish three different notions of contextuality:
\begin{itemize}
\item {\bf Bohr-contextuality:}  ``...the impossibility of any sharp separation between the behaviour of atomic objects and the interaction with the measuring instruments ...'' (\cite{BR0}, v. 2, p. 40-41; see also \cite{PL2,KHBb,NL2}).
\item {\bf Joint-measurement contextuality:} If $A, B, C$ are three quantum observables, such that $A$ is compatible with $B$ and  $C,$ a measurement of $A$ might give different result depending upon whether $A$ is measured  with $B$ or with $C.$ 
\item {\bf Bell-contextuality:} violation of noncontextual (Bell-type) inequalities.\footnote{This approach to contextuality was actively driven by Adan Cabello; so it may be natural to call it Bell-Cabello contextuality.} 
\end{itemize}

Bohr-contextuality is a part of the Bohr's complementarity principle \cite{BR0,PL2,KHBb,NL2} (see Appendix 1); 
it is closely connected  with incompatibility. Joint-measurement contextuality is the very special case of Bohr-contextuality. Bell-contextuality, at least for quantum physical observables, is a consequence of existence of incompatible observables, i.e., it is also 
a consequence of the complementarity principle.    

Bohr-contextuality is experimentally tested through incompatibility and theoretically it is formulated in terms of commutators.
The basic test is based on the Heisenberg uncertainty relation in its general form of the Schr\"odinger-Robertson inequality. 
The joint measurement contextuality could ``exist'' only for incompatible observables $B$ $C,$ i.e., $[B,C]\not=0.$
However, for such observables $B$ and $C,$  this sort of contextuality is  not testable  experimentally due to its counterfactual structure.\footnote{Many years ago, Svozil proposed to proceed in the counterfactual framework towards design of the real physical   experimental test, see \cite{KS1,KS3} for description of this test. Unfortunately, he did not elaborate this framework and these papers are practically forgotten.} Bell-contextuality can be tested experimentally in experiments by demonstration of violation of various Bell-type inequalities. 

In this paper,  we {\it compare Bell-contextuality and incompatibility} (and, hence, indirectly via incompatibility we compare Bell and Bohr contextualities). Since Bohr contextuality will be discussed only in Appendix 1, throughout the paper {\it we shall call Bell-contextuality simply contextuality.}

It is easy to show that, for quantum observables, there is {\it no contextuality without incompatibility:} for compatible observables,
it is impossible to violate any noncontextuality inequality (Theorem 1, section \ref{QMC}).
The natural question arises:  

\medskip

{\it Has contextuality without incompatibility any physical meaning?}

\medskip

Generally this is the very complex question. I do not know the answer to it for general noncontextuality inequalities. And I hope that 
this paper would stimulate foundational research in this direction.   We concentrate on contextuality for four quantum observables -
noncontextuality analog of the CHSH-inequality.  

We proved that, for ``natural quantum observables'' , {\it contextuality is reduced to incompatibility}\footnote{
Outside of physics, e.g., in cognitive  psychology and  decision making  contextuality distilled of incompatibility can play the  important role \cite{BC1}-\cite{BC6}.}  (in \cite{NL1}-\cite{NL3}, the same conclusion was obtained for quantum nonlocality, cf. 
\cite{K1}-\cite{PGF}).

At the same time, we  shown that generally contextuality without incompatibility may have  some physical 
content. We found a mathematical constraint extracting the contextuality component from incompatibility. However, the physical meaning of this constraint is not clear. 

We also remark that there exist positive answers to the inverse question: there can be (non-quantum, quasi-classical) incompatibility without contextuality; as exposed by finite automata \cite{MR} as well as for generalized urn models \cite{WR}, see \cite{KS}. 

\section{Quantum theory: Bell-contextuality vs. Bohr-incompatibility}
\label{QMC}

In this paper, we  consider dichotomous observables taking values $\pm 1.$ 

We follow paper \cite{AD} (one of the best and clearest representations of contextuality).
Consider a set of observables $\{X_1, . . . ,X_n \};$  a  context $C$ is  a  set  of  indexes  such that $X_i, X_j$ 
are compatible for all pairs $i,j \in C.$  A  contextuality  structure for these observables is given by  a  set 
 of contexts ${\cal C}=\{ C \},$ or simply the  maximal  contexts.    For  each  context $C,$  we measure pairwise 
correlations   for observables $X_i$ and $X_j$ with indexes $i, j \in C$ as well as averages 
$\langle X_i\rangle$ of observables $X_i.$ The $n$-cycle  contextuality  scenario  is  given  by $n$ observables $X_1, . . . ,X_{n}$  and the set of maximal contexts
\begin{equation}
\label{C}
{\cal C}_n=\{\{X_1,X_2\}, . . . ,\{X_{n-1},X_{n}\},\{X_{n},X_1\}\}.
\end{equation}
Statistical data associated with this set of contexts is given by the collection of averages and correlations:
\begin{equation}
\label{Cw}
\{\langle X_1\rangle,...., \langle X_n\rangle; \langle X_1 X_2\rangle, . . . , \langle X_{n-1}X_{n}\rangle, \langle X_{n},X_1\rangle\}.
\end{equation}

Theorem 1 from paper \cite{AD} describes all tight noncontextuality inequalities. In particular, for $n=4$ we have inequality:
\begin{equation}
\label{P7}
  \vert \langle X_1  X_2  \rangle +  \langle X_2 X_3  \rangle + \langle X_3 X_4 \rangle  - \langle X_4 X_1\rangle \vert \leq 2 .
\end{equation}

Theorem 2 \cite{AD} demonstrates that, for $n \geq 4$ (cf. appendix 1 for $n=3),$ aforementioned tight noncontexuality inequalities  are violated by quantum correlations. But,

\medskip

{\it  what is the physical  root of quantum violations?} 

\medskip

Unfortunately, the formal mathematical calculations \cite{AD} used to show violation 
of noncontextuality inequalities for quantum observables do not clarify physics behind these violations.   

Let us turn to the quantum physics, i.e., $X_1, . . . ,X_{n}$ are not arbitrary observables, but quantum physical ones. 
In the quantum formalism, they are represented by Hermitian operators $\hat X_1, . . . , \hat X_{n}.$ 
Denote the orthogonal projectors onto the corresponding eigenspaces by the symbols  $\hat E_{j \alpha}, \alpha = \pm 1.$ 

Suppose now that these  observables are compatible with each other, i.e., any two observables $X_i,X_j$ can be jointly measurable, 
so in the operator formalism, $[\hat X_i,\hat X_j]=0.$ The quantum theory has one amazing feature that is not so widely emphasized: 
\medskip

{\it Pairwise joint measurability  implies $k$-wise  joint measurability for any $k\leq n.$ }

\medskip

 If all pairs can be jointly measured, then even any family of observables
$\{ X_{i_1},..., X_{i_k} \}$ can be jointly measured as well. In principle, there is no reason for this. This is the specialty 
of quantum theory.  

The joint probability distribution (JPD) of compatible observables is defined by the following formula \cite{VN}:
\begin{equation}
\label{P}
P_{i_1...i_k} (\alpha_{i_1},...,\alpha_{i_k})
= \rm{Tr} \rho \hat E_{i_1 \alpha_{i_1}} \cdots \hat E_{i_{k} \alpha_{i_k}}.
\end{equation}
In particular, by setting $k=n$ we obtain JPD of all observables, 
\begin{equation}
\label{P1}
P_{1...n} (\alpha_{1},..., \alpha_{n})=\rm{Tr} \rho \hat E_{1 \alpha_{1}} \cdots \hat E_{n \alpha_{n}}.
\end{equation}
 We remark that the probability distributions given by (\ref{P}) can be 
obtained from the latter JPD as the marginal probability distributions:
\begin{equation}
\label{P3}
P_{i_1...i_k} (\alpha_{i_1},...,\alpha_{i_k})= \sum_{\alpha_j, j\not= i_1...i_k}
P_{1...n} (\alpha_{1},..., \alpha_{n}).
\end{equation}
This formula implies as well that the marginals of JPD  $P_{i_1...i_k}$ of the rank $k$ generate 
JPDs of the rank $k-1.$ In particular, we have the consistency rules for JPDs of ranks 2 and 1, 
\begin{equation}
\label{P4}
P_{i} (\alpha_i)= \sum_{\alpha_j}
P_{ij} (\alpha_i,\alpha_j)
\end{equation}
(in quantum physics, this condition is known as no signaling), and ranks 3 and 2 consistency:
\begin{equation}
\label{P4}
P_{ij} (\alpha_i, \alpha_j)= \sum_{\alpha_k}
P_{ijk} (\alpha_i,\alpha_{j}, \alpha_k)
\end{equation}
We have the classical probability framework; the Kolmogorov probability model with the probability measure $P\equiv P_{1...n}.$ In this classical probabilistic framework we can prove any noncontextuality  inequality (any Bell-type inequality, cf. \cite{AFine}-\cite{MMM}, 
\cite{KHBb,KHB2}). It is impossible to violate them for compatible quantum observables. We can formulate this result as a simple mathematical statement:

\medskip

{\bf Theorem 1.} {\it For quantum observables $X_1,...,X_n,$ (Bell-)contextuality implies 
incompatibility of at least two of them.} 

\medskip

Thus, there is no Bell-contextuality without incompatibility. Does the latter contain something more than incompatibility? 

Finally, we remark that noncontextuality inequalities started to be used in applications outside of physics, e.g., 
in psychology, cognitive science, and decision making \cite{BC1}-\cite{BC6}. If one does not assume that observables are represented by Hermitian
operators in Hilbert space, then ``no-go'' Theorem 1 loses its value.  

\section{Is contextuality reduced to incompatibility? }

In \cite{NL1}, I analyzed in details  the CHSH-inequality; the CHSH-correlation has the form:
\begin{equation}
\label{P5}
\Gamma =\langle A_1 B_1  \rangle + \langle A_1 B_2 \rangle + \langle A_2  B_1  \rangle- \langle A_2 B_2 \rangle, 
\end{equation}
where observables $A_i$ are compatible with observables $B_j, i,j=1,2.$  In \cite{NL1}, the tensor product structure of the state space was not explored and quantum observables were represented by Hermitian operators  $\hat A_i, \hat B_j$ acting an arbitrary Hilbert space. 
In this framework the CHSH-inequality can be treated as the noncontextuality inequality for four observables; by setting  in 
(\ref{P5}) $A_2=X_1, B_1=X_2, A_1=X_3, B_2=X_4,$ we obtain the correlation: 
\begin{equation}
\label{P5a}
\Gamma = \langle X_1  X_2  \rangle +  \langle X_2 X_3  \rangle + \langle X_3 X_4 \rangle  - \langle X_4 X_1 \rangle, 
\end{equation}
since we work with quantum observables, we proceed under the compatibility assumption
\begin{equation}
\label{P5b}
[\hat X_1, \hat X_2]=0, [\hat X_3, \hat X_2]=0, [\hat X_3, \hat X_4]=0, [\hat X_1, \hat X_4]=0.  
\end{equation}
Now set 
\begin{equation}
\label{P8}
\hat M_{13}= i [\hat X_1, \hat X_3] \; \mbox{and}  \; \hat M_{34}= i [\hat X_2, \hat X_4].
\end{equation}
These are Hermitian operators, so they represent some  quantum observables $M_{13}$ and $M_{34}.$ 
We remark that these observables are compatible:
\begin{equation}
\label{P8d}
[\hat M_{13}, \hat M_{34}] =0.
\end{equation}

The following theorem is the noncontextuality reinterpretation of the main result of paper \cite{NL1}:

\medskip

{\bf Theorem 2.} {\it Condition
\begin{equation}
\label{P6}
\hat M_{13} \circ \hat M_{34} \not= 0.
\end{equation}
 is necessary and sufficient  for violation of 
the noncontextuality inequality (\ref{P7})
for some quantum state.}

\medskip

{\bf Proof's scheme.} Consider the operator 
\begin{equation}
\label{P6P}
\hat \Gamma = \hat X_1 \hat X_2  +   \hat X_2 \hat X_3   +  \hat X_3 \hat X_4   - \hat X_4 \hat X_1. 
\end{equation}
Then we have
\begin{equation}
\label{P6P1}
\hat \Gamma^2 = 4 + [\hat X_1, \hat X_3] [\hat X_2, \hat X_4] = 4 + \hat M_{13} \hat M_{34}. 
\end{equation}
Then it is easy to show  that $\Vert \hat \Gamma^2 \Vert >4,$ if and only if condition (\ref{P6}) holds. 
Finally, we note that 
\begin{equation}
\label{P6P2}
\sup_{\Vert \psi\Vert=1} \vert \langle \psi\vert \hat \Gamma\vert \psi \rangle =   \Vert \hat \Gamma \Vert = 
\sqrt{\Vert \hat  \Gamma^2 \Vert}.  
\end{equation}

\medskip

We remark that condition (\ref{P6}) is trivially satisfied for incompatible observables,  if the state space and observables have the tensor product structure:
$H=H_{13} \otimes H_{24}$ and 
\begin{equation}
\label{P8}
\hat X_i={\bf \hat X}_i \otimes I,   \hat X_j=I\otimes {\bf \hat X}_j, 
\end{equation}
where 
\begin{equation}
\label{P8n}
{\bf \hat X}_i : H_{13} \to H_{13}, i=1,3,   {\bf \hat X}_j: H_{24} \to H_{44}, j=2,4.
\end{equation}
Here condition (\ref{P6})  is reduced to incompatibility condition:
\begin{equation}
\label{P8m}
[{\bf \hat X}_i, {\bf :\hat X}_j] \not= 0, i=1,3; j=2,4.
\end{equation}
In particular, for compound systems, contextuality (``nonlocality'')  is exactly incompatibility. The same is valid for any tensor decomposition of the state space of a single quantum system with observables of the type (\ref{P8}). In the tensor product case, 
contextuality without incompatibility leads to the notion with the empty content. 

But, it may happen that $X_i$-observables, $i=1,3,$ and $X_j$-observables, $j=2,4,$ are not connected via the tensor product structure. In this case, the interpretation of constraint (\ref{P6}) is nontrivial. What is its physical meaning? I have no idea.

Of course, the main problem is that it is not clear at all how to measure  the observables of the commutator-type.

\medskip

\section{Conclusion} 

In quantum physics, there is no contextuality without incompatibility. This is well known, but not so highly emphasized 
feature of quantum observables. 

For fourth quantum observables, these 
two notions coincide under validity of constraint (\ref{P6}). If it is violated,  then, for such observables,  there is still a hope that quantum contextuality without incompatibility has some nontrivial physical meaning. (What?)
The problem of nontrivial physical meaning of ``pure contextuality'', i.e., one distilled from incompatibility, for $n>4$  observables (as well as $n=3,$ see appendix 2) is open. 

Finding the right physical interpretation for contextuality beyond incompatibility is important for demystification of quantum physics
(cf. with discussion of Svozil \cite{KS} on ``quantum focus pocus''). 
  	
\section*{Appendix 1: Structuring Bohr's contextuality and complementarity into a single principle}

As was emphasized in \cite{KHB2}, the complementarity principle is closely coupled with the notion of contextuality that is understood in Bohr's sense. 
Bohr did not use  the notion  ``experimental context''. He  operated with the notion of  experimental
condition \cite{BR0}:

\medskip

{\it ``Strictly speaking, the mathematical formalism of quantum mechanics
and electrodynamics merely offers rules of calculation for the deduction of
expectations pertaining to observations obtained under well-deffined experimental 
conditions specified by classical physical concepts.''}

\medskip

Unfortunately, Bohr did not formulate his views on quantum foundations in the form 
of principles, similar to Einstein's principles of relativity. These views were presented in the form of the foundational 
statements connected with long texts on the general structure of quantum theory and its methodology, especially methodology of 
quantum measurements. And these statements were often modified year to year. Nevertheless, careful reading of Bohr's works leads
to clear picture of quantum foundations. We remark that in this picture there is nothing mystical or too much surprising. This 
is logically well structured reasoning on specialty of quantum measurements (and, for Bohr, the quantum theory is a measurement theory). In my previous papers, I called this bunch of Bohr's views the complementarity principle. This can lead to misunderstanding. Nowadays, the 
complementarity principle is typically reduced to the wave-particle duality - the existence of incompatible observables (experimental contexts). 
The latter is just the concluding accord of long Bohr's play, the play on contextuality of quantum measurements. Since this paper 
is devoted to contextuality, this is the good place to restructure my formulation \cite{KHB2,NL1,NL3} of the Bohr's complementarity 
principle \cite{BR0} - to highlight its contextual counterpart.  

We start with pointing to the physical basis of quantum contextuality and complementarity. 
Bohr stressed \cite{BR1,BR2a} that the essence of the quantum theory  {\it ``may be expressed in the so-called quantum postulate, which attributes to any atomic process an essential discontinuity, or rather individuality, completely foreign to the classical theories and symbolised by Planck's quantum of action.''} This postulate is about nature as it is. And the postulate is the root of the fundamental principles of the quantum theory (the quantum measurement theory).

We continue with the famous citation of Bohr that presents the essence of his views  on contextuality and complementarity of quantum measurements, see Bohr (\cite{BR0}, v. 2, p. 40-41): 

\medskip

``This crucial point ...  implies {\it the impossibility of any sharp separation between the behaviour of atomic objects and the interaction with the measuring instruments which serve to define the conditions under which the phenomena appear.} In fact, the individuality of the typical quantum effects finds its proper expression in the circumstance that any attempt of subdividing the phenomena will demand a change in the experimental arrangement introducing new possibilities of interaction between objects and measuring instruments which in principle cannot be controlled. Consequently, evidence obtained under different experimental conditions cannot be comprehended within a single picture, but must be regarded as complementary in the sense that only the totality of the phenomena exhausts the possible information about the objects.''
     
\medskip

By the quantum postulate there exists indivisible quantum of action  given by
the Planck constant $h.$ Its presence  prevents approaching the internal features
of a quantum system.  Therefore it is meaningless (from the viewpoint of physics)
to build scientific theories about such features. This reasoning (rooted in the quantum 
postulate) implies:

 \medskip

{\bf Principle of Contextuality:} {\it  The output of any quantum observable is indivisibly composed of the contributions of the system and the measurement apparatus.}

\medskip

There is no reason to expect that all experimental contexts
can be combined and all observables can be measured jointly. Hence, 
incompatible observables (complementary experimental contexts) may exist. Moreover, they should exist, otherwise the 
contextuality principle would have the empty content.  Really, if all experimental
contexts can be combined into single context ${\cal C}$ and all observables can be jointly measured in this context, then the outputs of such joint measurements can be assigned directly to a system. To be more careful, we have to say: ``assigned to a system and context ${\cal C}''.$ But, the latter can be omitted, since this is the same context for all observables. This reasoning implies:         

\medskip

{\bf Principle of Complementarity:} {\it There exist incompatible observables (complementary experimental contexts).}

\medskip

Since both principles, contextuality and complementarity, are so closely interrelated, it is natural to unify them into the single
principle, {\bf Contextuality-Complementarity principle.} 

Bohr's viewpoint on contextuality and its coupling with complementarity  was explored in a series of author's papers, see, e.g., monograph  \cite{KHBb}. 

This is the right place to stress once again the difference between {\it Bohr-contextuality} and the notion of contextuality that 
is widely used in considerations related to the Bell-type inequalities, {\it Bell-contextuality} 
\cite{Bell1,Bell2}.\footnote{We remark that Bell did not use the term ``contextuality''. } The former has no relation to joint measurement. It is about context-dependence of outputs of a single observable. Of course, joint measurement of a compatible observable can also be considered as specification of experimental context. However, such viewpoint on contextualization only overshadow the original Bohr's view: contextuality as impossibility to separate (in measurement's output) the contributions of the system and  measurement device.   

We remark that coupling of  the contextuality principle to the quantum postulate, the existence of the Planck constant, is important only for foundations of  {\it quantum physics.} Outside of physics, one can start directly with  the contextuality principle. It can be applied even to nonphysical systems, see, e.g., \cite{BC1}-\cite{BC6} on applications to decision making, cognitive and social sciences. However, the class of  observables described by quantum mechanics is very special; they are represented by Hermitian operators acting in compelx Hilbert space, see section  \ref{QMC} for foundational consequences.

In the line of the above reasoning (I hope that Bohr would agree with it), the existence of complementary experimental contexts and incompatible observables is very natural, there is nothing mystical in this. This is a consequence of (Bohr-)contextuality and the latter in turn is a consequence of the quantum postulate. Since we claim that quantum nonocality
and Bell-contextuality are reduced to the existence of incompatible observables - the principle of complementarity, it seems that the only mystery of quantum physics is the quantum postulate (see \cite{H} for details). 
        	
\section*{Appendix 2. Suppes-Zanotti inequality: Has it any relation to quantum physics?}

The case $n=3, X_1, X_2, X_3,$ is special. Here the tight noncontextuality inequality  was derived by 
Suppes and Zanotti \cite{SZ}: 
\begin{equation}
\label{SZ}
\langle X_1 X_2 \rangle - \langle X_2 X_3 \rangle + \langle X_1 X_3 \rangle \leq 1.
\end{equation}
Often this inequality is misleadingly coupled to the original Bell inequality. However,   the  Suppes-Zanotti
inequality has nothing to do with quantum mechanics. Since it is assumed that all pairs of observables are compatible,   
the JPD for quantum observables always exists and this inequality is always satisfied. So, the criterion of the existence 
of  JPD derived in \cite{SZ} has no relation to quantum mechanics. 

The {\it original Bell inequality} has the form: 
\begin{equation}
\label{SZ1}
\langle X_1 X_2 \rangle - \langle X_3  X_4 \rangle + \langle X_1  X_4 \rangle \leq 1.
\end{equation}
Here observable $X_1$ should be compatible with observables $X_2, X_4$ and $X_3$ with $X_4.$  
The crucial condition for its derivation is  the precise correlation condition 
\begin{equation}
\label{SZ2a}
<X_2, X_3> = 1 .
\end{equation}
(Unfortunately, I simply forgot to mention this condition in paper \cite{NCONT}.)  This is the inequality based  on three contexts for four observables. It is not a tight noncontextuality inequality, so it is not covered by Theorem 1 \cite{AD}.  
Surprisingly this inequality  is more complicated than inequality (\ref{P7}), see \cite{EL} for some steps towards its analysis.

This inequality differs crucially from other Bell-type inequalities, because of the correlation constraint (\ref{SZ2a}). 
Its experimental violation should be checked for quantum states with perfect correlations for at least one pair of observables.
On the other hand, only this inequality has coupling with  the original EPR argument, precisely because this anti-correlation condition. The CHSH-inequality \cite{CHSH} has not so much to do with EPR-paper \cite{EPR} (in spite of the rather common opinion that 
by demonstrating its violation, it was demonstrated that ``Einstein was wrong'').  The CHSH-inequality was tested just because it was easier to test  (it is difficult to perform experiments for highly entangled states, see article \cite{OBELL} for detailed analysis).\footnote{It is easier to search for  the lost wallet under a streetlight than int he  darkness. And after years of search, one can be happy by finding something reminding the wallet, although not the original one.}

We remark that the Suppes-Zanotti inequality coincides with the inequality for three random variables $(X_1,X_2,X_3)$ 
derived by Boole as a necessary condition of existence of their JPD $p_{X_1 X_2 X_3}.$ This ``discovery'' was made by 
I. Pitowsky. However, he identified the Suppes-Zanotti inequality with the original Bell inequality. This statement\footnote{At least, Pitwosky represented the situation in this way at the V\"axj\"o conference ``Quantum Theory: Reconsideration of foundations'' in 2001.}
spread widely in the form that Bell-inequality was well know even to Boole and I am also responsible for this  \cite{KC5}.\footnote{I told about this Boole-Bell coupling to K. Hess and W. Pilipp and then they spread such a viewpoint further.}.

The common point in Boole and Bell reasoning is the use of JPD $p_{X_1 X_2 X_3}$ (although Bell did not highlight this crucial point). However, Bell's main aim was to proceed with the EPR-argument, based on the use of perfect correlations. This point was not present in Boole's considerations that were directed solely to the JPD-existing problem.

 \end{document}